\documentclass[aps,prb,twocolumn,groupedaddress,showpacs,floatfix]{revtex4}

\usepackage{graphicx} \usepackage{epsfig} \usepackage{dcolumn}
\usepackage{bm} \usepackage{amsmath} \usepackage{verbatim}
\usepackage[usenames]{color} 

\newcommand{\figa} {
\begin{figure}[t]
\centering \includegraphics[width=0.45\textwidth]{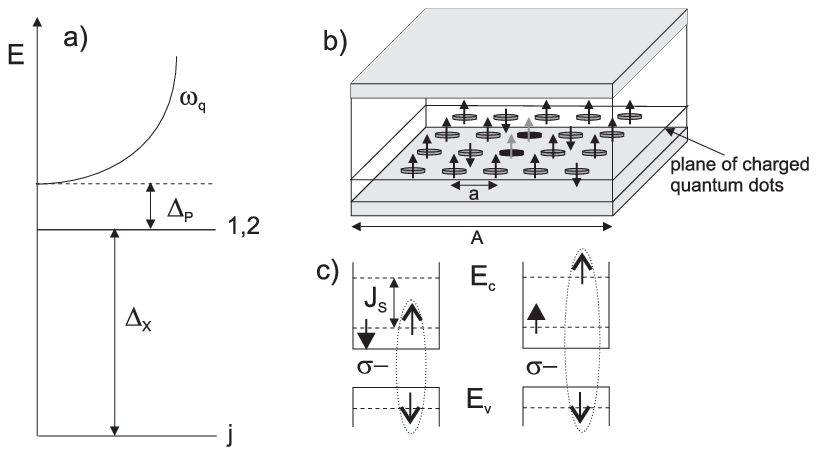}
\caption{(a) Energy diagram with cavity photon dispersion, and detunig
$\Delta_P$ and $\Delta_X$ discussed in the text. (b) Scheme of the
quantum memory composed by charged quantum dots in a planar
cavity. Two dots brought into resonance with the cavity are
highlighted. (c) Diagram of allowed spin configurations for a charged
dot excited by circulartly polarized light. The distance between the
trion energy in the two configuration defines an antiferromagnetic spin
coupling between the electron spin and the exciton spin
(polarization).}
\label{figa}
\end{figure}
}

\newcommand{\figab} {
\begin{figure}[t]
\centering \includegraphics[width=0.25\textwidth]{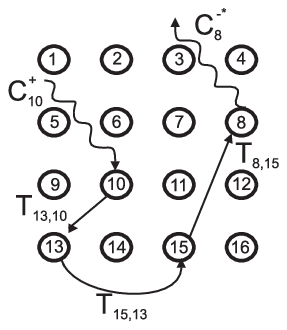}
\caption{Diagram illustrating multiple scattering events that lead to
a multi-spin coupling $J^{(4)}_{8,15,13,10}$, as derived in
Eq. (\ref{couplpertur}).}
\label{figab}
\end{figure}
}

\newcommand{\figb} {
\begin{figure}[t]
\centering
\includegraphics[width=0.4\textwidth]{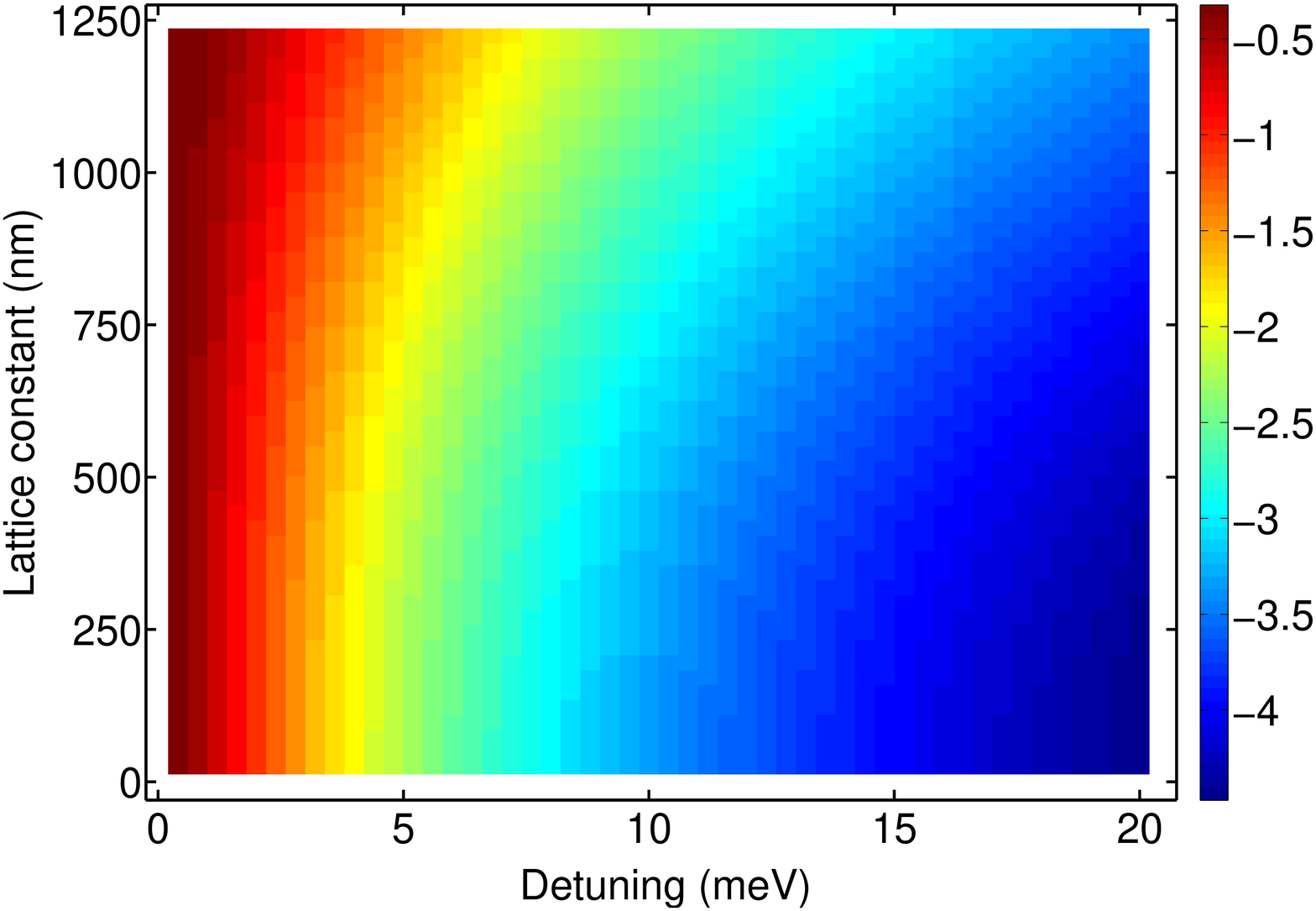}
\caption{(Color online) Logarithmic plot of the error $\mathcal{E}$ as a function of
the detuning $\Delta_X$ and lattice constant $a$ in a Phase Gate
between two most distant dots in a $3 \times 3$ array of charged
QDs,$\Delta_P = 1\,$meV.}
\label{figb}
\end{figure}
}

\newcommand{\figd} {
\begin{figure}[t]
\centering
\includegraphics[width=0.4\textwidth]{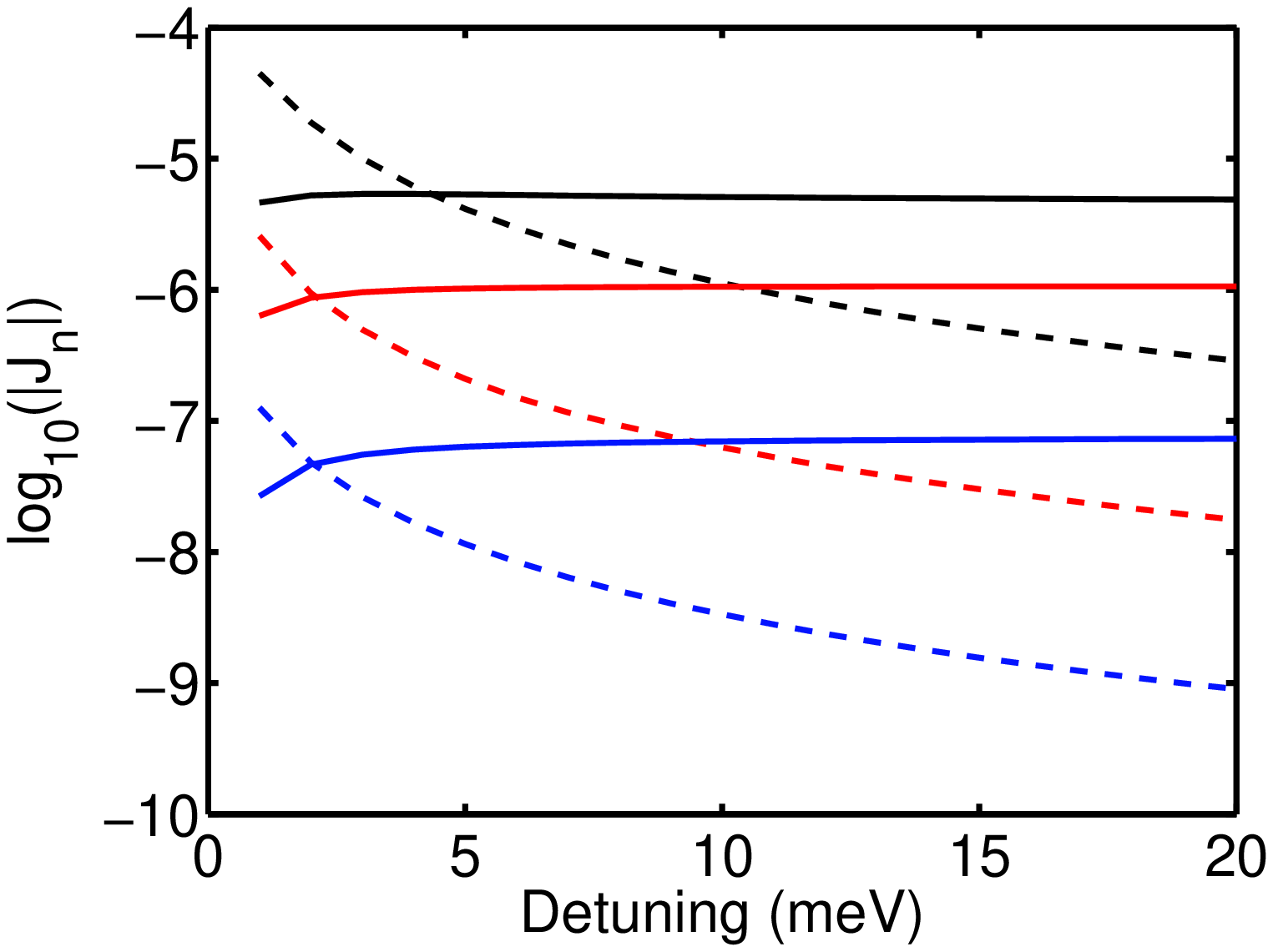} 
\caption{(Color online) The logarithmic plot of $J^{(n)}_R$ (solid) and $J^{(n)}_O$ (dashed) [see text for details] as a function of the detuning $\Delta_X$ in a $3 \times 3$ array of charged QDs with $\Delta_P = 1\,$meV and the lattice constants $a = 100\,$nm for $n=2$ (black), $n=4$ (red), and $n=6$ (blue) are shown.
}
\label{figd}
\end{figure}
}

\begin{document}
\title{Multi-spin errors in the optical control of a spin quantum
memory}

\author{Michal Grochol and Carlo Piermarocchi}

\affiliation{Department of Physics and Astronomy, Michigan State
University, East Lansing, Michigan 48824 USA}

\date{\today}
\begin{abstract}
We study a quantum memory composed of an array of charged quantum dots
embedded in a planar cavity. Optically excited polaritons,
i.e. exciton-cavity mixed states, interact with the electron spins in
the dots. Linearly polarized excitation induces two-spin and
multi-spin interactions. We discuss how the multi-spin interaction
terms, which represent a source of errors for two-qubit quantum gates,
can be suppressed using local control of the exciton energy. We show
that using detuning conditional phase shift gates with high fidelity
can be obtained. The cavity provides long-range spin coupling
and the resulting gate operation time is shorter than the spin decoherence
time. 
\end{abstract}


\maketitle

\paragraph{Introduction}

In the last few years there have been great advances towards quantum
information processing in the solid state.  Yet, there are many
theoretical and practical problems that remain to be addressed. In
particular, there is not yet a solid state systems for which all the
feasibility criteria for quantum computing (i.e. decoherence, reliable
one- and two-qubit operations, scalable qubit, initialization and
read-out~\cite{DiVin2000}) have been simultaneously
demonstrated. Lately, electron spins in semiconductors, localized
either in low-dimensional nanostructures, i.e. QDs or in impurities,
are increasingly receiving attention as qubits due to their very long
decoherence time, which is typically of the order of $T_2 =3 \mu$s.
\cite{KDH+2004,PJT+2005,GYS+2006} The long coherence time of the electron spin
is due to its weak interaction with the environment, which on the
other hand makes its control more demanding. In this framework,
optical techniques are very promising since in this case the control
is realized using an optically active ancillary excited state, e.g. a
trion state in quantum dots, leading to a control that can be obtained
in picoseconds. Optical initialization,
\cite{DCL+2005,EXS+2007,XWS+2007} single qubit measurement,
\cite{CKL+2002,ADB+2007} and selective one-qubit control of QD's spin
\cite{WKX+2007,BMS+2008} have been already
demonstrated. Similar experiments on impurity states have also been carried out.\cite{FYC+2006,STN+2006}
The two-qubit control represents a more challenging
task. Optically mediated long range spin-spin interaction in a cavity
system has been explored theoreticaly only for two QDs.
\cite{IAB+1999,QFP2006}

In this paper, we show that an array of charged QDs embedded in a
planar cavity (see Fig. \ref{figa}) is a good candidate for a
controlable quantum memory. We extend the previous works on polariton
mediated spin coupling~\cite{QFP2006} to the case of many dots, which
leads to the appeareance of {\it multi-spin} Ising-like coupling
terms. We consider a system in which the energy of the ancillary
states on each dot can be controlled independently, for instance using
gates on each dot. We calculate the fidelity of the phase gate of two
spins being in resonance with the cavity mode and show that by
controlling the detuning of the remaining dots, gates with very small
error can be obtained. Errors due to multi-spin terms in the case of
quantum dot directly coupled by wavefunction overlap have also been
studied recently.\cite{ML2004} The model of multi-spin coupling is
also applicable to similar systems like e.g. superconducting qubits
embedded in a cavity, for which the two-qubit control has been
demonstrated in a recent experiment.\cite{MCG+2007}

\paragraph{Polariton-Spin Hamiltonian}

\figa 

Our assumptions for the system studied are the following: (i) the
trion energy $\Delta_{X,j}$ of each dot can be independently
controlled e.g. by applying a local voltage,\cite{RRL+2005} (ii) the
quantum dots are well separated so there is not direct overlap of the
trion wavefunction, (iii) each dot can be occupied only by one
additional exciton, (iv) the heavy-hole light-hole splitting is large
enough that only the heavy-hole exciton is taken into account, and (v)
the cavity is ideal. The role of the cavity is to enhance the range of
the interaction between dots\cite{tarel07} and their spins.
\cite{QFP2006} The Hamiltonian descibing the memeory can therefore be
written as ($\hbar = 1$ throughout the paper)
\begin{eqnarray}
\label{hamil1}
\nonumber \hat{H}_g &=& \sum_\alpha \Bigl \{ - \sum_{j} \Delta_{X,j}
 C^\dagger_{j \alpha} C_{j \alpha} \\ &+& \sum_{q j} (g_{q j} a_{q
 \alpha} C^\dagger_{j \alpha} + h.c.) + \sum_{q} \omega_q
 a^\dagger_{\alpha q} a_{\alpha q} \Bigr \} \\ \nonumber &+& \sum_j
 J_S S_{j z} P_{jz} + \hat{H}_L,
\end{eqnarray}
where $C^\dagger_{j \alpha}$ ($C_{j \alpha}$) is the creation
(annihilation) operator of exciton on the $j^{th}$ dot at position
$R_j$ with polarization $\alpha$, $a^\dagger_{\alpha q}$ ($a_{\alpha
q}$) is the creation (annihilation) operator of the photon with
two-dimensional momentum $q$, $g_{q j} = g \, e^{-q^2 \beta^2}
e^{iqR_j}$ is the dot-photon coupling constant with $\beta$ being the
effective dot size, $J_S$ is the energy difference between trion states with
parallel and anti-parallel spins as schematically shown in
Fig.~\ref{figa}c, $S_{j z}$ is the $z$-component of the electron spin
in the $j^{th}$ QD, and $P_{jz} = C^\dagger_{j \uparrow} C_{j
\uparrow} - C^\dagger_{j \downarrow} C_{j \downarrow}$ is the operator
corresponding to the $z$ component of the exciton polarization.  A
$\sigma+$ ($\sigma-$) polarized photon creates a bright exciton with
$\downarrow$ ($\uparrow$) electron spin in the growth ($z$)
direction. For excitons in III-V confinded systems the
possible values of the electron spin are $\sigma^e_z = \pm
\frac{1}{2}$ and the heavy hole spin are $\sigma^{hh}_z = \mp \frac{3}{2}$. We
assume troughout the paper that the light is linearly polarized. This
choice simplifies considerably the multi-spin problems since it makes
all multi-spin terms of odd order identically zero. The coupling of the cavity to the external electromagnetic field is
described using the quasi-mode model\cite{savona99} $\hat{H}_L =
\sum_{\alpha q} (V_{\alpha q} e^{i \omega_L t} a_{\alpha q} + h.c.)$,
where $V_{\alpha q}$ is the laser-cavity coupling constant
proportional to the cavity area $\sim \sqrt{A}$.

\paragraph{Multi-Spin Hamiltonian}
The effective spin Hamiltonian can be calculated introducing the level
shift operator $R(\omega_L)$ as \cite{Cohen2004} 
\begin{eqnarray}
\hat{H}_{s} =
\mathcal{P} R(\omega_L) \mathcal{P} = \mathcal{P} \hat{H}_L
\frac{\mathcal{Q}}{\omega_L - \mathcal{Q} \hat{H} \mathcal{Q}}
\hat{H}_L \mathcal{P}, 
\end{eqnarray}
where $\mathcal{P} = \sum_\lambda |\lambda
\rangle \langle \lambda | \otimes | 0 \rangle \langle 0 |$ $\Bigl [
\mathcal{Q} = 1 - \mathcal{P} = \sum_{\lambda \beta} |\lambda \rangle
\langle \lambda | \otimes | \beta \rangle \langle \beta | \Bigr ]$ is
the projection operator on the subspace of all spin states $\lambda$
and zero [one] excitation. Assuming the rotating wave approximation
and linearly polarized laser light propagating perpendicularly to the
cavity plane ($q = 0$) then $\hat{H}_L = V_{\downarrow 0} a_{0
\downarrow} + V_{\uparrow 0} a_{0 \uparrow} + h.c.$.

\figab

By solving first the polariton problem for $q = 0$ and both
polarizations, we can write $\hat{H}_P | \alpha \uparrow (\downarrow) \rangle =
\omega_\alpha | \alpha \uparrow (\downarrow) \rangle $ and $| \alpha
\uparrow (\downarrow) \rangle = \bigr ( \sum_j u_{\alpha j}
C^\dagger_{0 j \uparrow (\downarrow)} + \sum_{\mathsf{Q}} v_{\alpha
\mathsf{Q}} a^\dagger_{0 \uparrow (\downarrow)} \bigl ) | 0 \rangle $ where $\mathsf{Q}$ is a reciprocal lattice vector of the dot lattice.
Then the matrix element between the spin state reads
\begin{eqnarray}
\label{spinen}
R_{\lambda \lambda'}= \sum_{\alpha \beta} v_{\alpha 0} v^*_{\beta 0 }
\sum_{\gamma = \uparrow, \downarrow} \frac{V^2_{\gamma 0}}{2} \langle
\alpha \gamma | \langle \lambda |(\omega_L - \hat{H})^{-1}| \lambda'
\rangle |\beta \gamma \rangle.
\end{eqnarray}
The off-diagonal terms $\langle \lambda |(\omega_L - \hat{H})^{-1}|
\lambda' \rangle$ are zero since all spin dependent terms are
proportional to $S_z$. This allows us to calculate the energies in 
Eq. (\ref{spinen}) exactly. Perturbation theory can also be applied
and, assuming linearly polarized light, only even contributions
($\sim J_S^{(2n)}$) are nonzero giving
\begin{eqnarray}
\hat{H}_T = \sum_{i > j} \tilde{J}^{(2)}_{ij} S_{i z} S_{j z} +
\sum_{i > j > k > l} \tilde{J}^{(4)}_{ijkl} S_{i z} S_{j z} S_{k z}
S_{l z} + \dots,
\end{eqnarray}
where the coupling constants are renormalized to take into account
multiple scattering, e.g.
\begin{eqnarray}
\tilde{J}^{(2)}_{12} = J^{(2)}_{12} + J^{(2)}_{21} + \sum_{i
\mathsf{P}} J^{(4)}_{\mathsf{P}(12ii)} + \sum_{ij \mathsf{P}}
J^{(6)}_{\mathsf{P}(12iijj)} + \cdots,
\end{eqnarray}
where $\mathsf{P}$ indicates a permutation of all the indices. The
$z$-coupling constants can be explicitely expressed as
\begin{eqnarray}
\label{couplpertur}
J^{(n)}_{i_1 ... i_n} = J^n_S V^2_{LP} \, (\mathsf{C}^{-}_{i_1})^*
\mathsf{T}_{i_1 i_2} \cdots \mathsf{T}_{i_{n - 1} i_n}
\mathsf{C}^+_{i_n}
\end{eqnarray}
in terms of the photon-exciton coupling function and exciton inter-dot
transfer probability (see scheme in Fig. \ref{figab})
\begin{eqnarray}
\label{transfer}
\mathsf{C}^{+ (-)}_{i } = \sum_\alpha \frac{v_{\alpha 0} u^*_{\alpha
i}}{\omega_L - \omega_\alpha \pm i \eta}, \; \mathsf{T}_{i j} =
\sum_\alpha \frac{u_{\alpha i} u^*_{\alpha j}}{\omega_L -
\omega_\alpha + i \eta},
\end{eqnarray}
where $\eta$ is the exciton and photon homogeneous broadening, assumed
identical for simplicity, and $V^2_{LP} = \frac{V^2_{\uparrow 0} +
V^2_{\downarrow 0}}{2}$ is the effective light-polariton coupling
constant.

Let us now consider two dots labeled by $\{ 1, 2 \}$ with a small
detuning with respect to the lowest cavity mode, i.e. $\Delta_{X,
1(2)} = \Delta_P$. The remaining quantum dots are detuned by a lager
amount: $\Delta_{X, j \ne 1,2} > \Delta_P$, as schematically plotted
in Fig. \ref{figa}~(a). Dots shifted off-resonance by a DC Stark shift will also have a weaker light-dot coupling $g$ due to the decrease of the electron-hole overlap. However, in order to have a conservative estimate of the error we neglect this effect.

We have used the following parameters: $\beta = 35\,$nm, $g = 70 \, \mu$eV, $\eta = 50
\, \mu$eV, $V_{\uparrow 0} = V_{\downarrow 0} = 0.9 \,$meV, $\omega_L
= \omega_{q=0}$, $J_S = 0.39$~meV, and exciton detuning up to
$\Delta_X = 20\,$meV, which is about the upper limit for a Stark shift
that can be obtained in current experiments. In the numerical calculation we
consider a finite system with 9 dots and we used periodic boudary
conditions in order to match the excitonic states in the dots with the
continuous two dimensional photon modes.

\figd 

The dependence of different multi-spin terms on the detuning is shown in Fig. \ref{figd} where we separate the terms that involve the two dots nearly resonant with the cavity from the others. We plot $J_{12}+J_{21}$ (solid black) and $\sum_{ij \notin \{1,2 \}} |J_{ij}|$ (black dashed) for $n=2$ spin terms. The contributions, that renormalize the effective coupling between $1$ and $2$ ($J_R^{(n)}$) from contributions that involve only the dots strongly detuned from the cavity ($J_O^{(n)}$), are separated for multi-spin terms ($n=4$, $n=6$). For instance, for $n=4$ the resonant (off-resonant) terms are defined as $J^{(4)}_R = \sum_{i \mathsf{P}} |J^{(4)}_{\mathsf{P}(12ii)}|$ ($J^{(4)}_O = \sum_{ijkl}  |J^{(4)}_{ijkl}| - |J^{(4)}_R|$). This definition enables us to better estimate the contribution of the off-resonant terms. In fact, even if the magnitude of the indivual terms $J^{(n)}_{i_1..i_n}$ is very small (e.g. $10^{-15}$ for $J^{(6)}_{ijklmn}$) we get a sizeable effect due to the large number of $n$-dot combinations ($\sim {n \choose N_D }$). Note that there is almost no dependence on the detuning for the resonant terms and a strong decrease for the off-resonant terms ($J^{(n)} \sim \Delta^{-(n - 1)}_X$) as expected from the form of the coupling in Eq. (\ref{couplpertur}).

\paragraph{Fidelity of a conditional phase shift gate}
We will now explicitely estimate the error in the implementation of a
conditional phase shift gate due to multi-spin interaction terms.  The
conditional phase gate (PG), is a universal two-qubit gate, i.e. can
realize universal quantum computation when combined with single qubit
operations.~\cite{BBC+1995} In the computational basis $\{|\downarrow
\uparrow \rangle,|\downarrow \uparrow \rangle,|\uparrow \downarrow
\rangle,| \uparrow \uparrow \rangle \}$ the PG can be written as
diagonal matrix with elements $U_{PG} = \{1,1,1,-1 \}$.  Assuming the
Ising-like interaction between two spins $\sim S_{z1} S_{z2}$, the
following sequence \cite{LBK+1999} gives the PG $U_{PG} = e^{i \pi /
4} [S_{z1} + S_{z2}] [ -2 S_{z1} S_{z2}]$ with $[P] = e^{i \pi / 2
P}$. A quantitative measure of the gate quality can be been given
using the gate \textit{fidelity}~\cite{PCZ1997} defined as 
$\mathcal{F} = \overline{|\langle \Psi |U_I^\dagger U_R| \Psi
\rangle|^2}$, where $U_I$ is the ideal gate matrix, and $U_R$ is the
real gate matrix, i.e. the one that includes the effects of multi-spin
terms. $\Psi$ is an arbitrary initial pure state, and
$\overline{|\langle \Psi |.| \Psi \rangle|^2}$ indicates averaging
over all pure states.  Working in the basis of the full
spin-Hamiltonian eigenstates $\{ \phi_i \}$ (with $2^{N_D}$ states),
we can define an eigenvector fidelity as $\mathcal{F}_i = \langle
\phi_i |U_I^\dagger U_R| \phi_i \rangle$. Since the total Hamiltonian
does not allow for spin flip processes, the fidelity can the be
expressed as $\mathcal{F} = |\frac{1}{N_D} \sum_i
\mathcal{F}_i|^2$. In order to calculate the fidelity, we calculate
the dynamics $\exp(-i H_R t_C)$, where the time $t_C$ is optimized so
to obtain maximal fidelity. The gate can be described as follows: (i)
two selected dots $\{ 1,2 \}$ are brought adiabatically into resonace
with the cavity by controlling the exciton energy with local electric
field, (ii) the laser is switched on for a time $t_C$, and (iii) dots
are brought back into the off-resonant state.

\figb
The calculated error $\mathcal{E}=1-\mathcal{F}$ as a function of the
detuning and lattice constant of the dot array is shown in
Fig. \ref{figb}. The fidelity $\mathcal{F}$ increases at larger
detuning $\Delta_X$ since only the two selected dots $\{1,2 \}$ remain
in resonance with the cavity and the multi-spin coupling with the other
dots is suppressed. On the other hand, increasing the lattice constant
decreases the fidelity since the exciton transfer, even if
considerably enhanced by the cavity, decreases with distance.
\cite{tarel07} The strong dependence of the fidelity on the detuning reflects the competition between the resonant and off-resonant terms as shown in Fig. \ref{figd}. Furthermore, note that the maximal value of the
individual dot detuning is limited by the inter dot separation  $a$. Then the
fidelity function $\mathcal{F}(\Delta,a)$ can be used  to select an optimal
lattice constant. 

Another important characteristic of the PG is its operation time,
i.e. the time during which the spin-interaction is switched on. The
operation time increases with increasing detuning since the spin-spin
coupling $\sim J_{12}$ decreases. Note that the time $t_C$ grows like
$\Delta_P^3$, following the dependence of the resonant terms in
Eqs. (\ref{couplpertur}) and (\ref{transfer}). Typical values of the
operation times are $t_C = 100\,$ps ($t_C = 450\,$ps) for $a =
100\,$nm ($a = 1300\,$nm) [$\Delta_X = 20\,$meV]. These characteristic
gate times are shorter than the spin decoherence time $T_2$, which is
of orders of at least $\mu s$.

\paragraph{Conclusions}
We have studied an array of charged quantum dots embedded in a planar
cavity as a candidate for the realization of a spin quantum memory. We
have shown that optical excitation can be used to control the spins
and implement quantum gates. The optical excitation couples many dots
in the quantum memory, and multi-spin interaction terms beyond the
ideal two-spin interaction are generated. We have shown that the
multi-spin terms can induce errors in the gate operation even if their
value is small, due to their multiplicity. These error can be
corrected by a local control of the excitonic resonance on each
dot. In the control scheme we also include a planar cavity that
modifies the photon density of states by providing a spectral region
where dots do not couple to radiation. The present control scheme can be applied to other similar solid state systems
like e.g., superconducting qubits embedded in a cavity.

This research was supported by the National Science Foundation, Grant
No. DMR-0608501.

\bibliographystyle{apsrev}

\end{document}